\documentclass[namedreferences]{SolarPhysics}
\usepackage[optionalrh]{spr-sola-addons} 
\usepackage{graphicx}                    
\usepackage{color}                       
\usepackage{url}                         
\usepackage{wasysym}

\newcommand{\etal}{{\it et al.}}

\begin{document}

\begin{article}

\begin{opening}

\title{STEREO-\textit{Wind} Radio Positioning of an Unusually Slow Drifting Event}

%
\author{J.C.~\surname{Mart\'{\i}nez-Oliveros}$^{1}$ \sep
        C.~\surname{Raftery}$^{1}$ \sep
        H.~\surname{Bain}$^{1}$ \sep
        Y.~\surname{Liu}$^{1,7}$ \sep
        M.~\surname{Pulupa}$^{1}$ \sep
        P.~\surname{Saint-Hilaire}$^{1}$\sep
        P.~\surname{Higgins}$^{6}$ \sep
        V.~\surname{Krupar}$^{2}$ \sep
        S\"am~\surname{Krucker}$^{1,4}$ \sep
        S.D.~\surname{Bale}$^{1,5}$ 
       }

%
\runningauthor{Mart\'{\i}nez-Oliveros \etal}
\runningtitle{STEREO-\textit{Wind} Radio Positioning of an Unusually Slow Drifting Event}

%
  \institute{$^{1}$ J.C.~Mart\'inez-Oliveros  \sep C.~Raftery \sep H.~Bain  \sep Y.~Liu \sep  M. Pulupa \sep S.~Krucker \sep S.D.~Bale \ \\ Space Science Laboratory, University of California, Berkeley, CA 94720, USA\\
                   email: \url{oliveros@ssl.berkeley.edu}, \url{claire@ssl.berkeley.edu}, \url{liuxying@ssl.berkeley.edu}, \url{pulupa@ssl.berkeley.edu}, \url{krucker@ssl.berkeley.edu}, \url{bale@ssl.berkeley.edu}\\
                   $^{2}$ V.~ Krupar \\ Institute of Atmospheric Physics ASCR, Bocni II 1403,Prague 141 31, Czech Republic\\ 
                   email: \url{vk@ufa.cas.cz}\\
		   $^{4}$ S\"am Krucker\\Institute of 4D Technologies, School of Engineering, University of Applied Sciences North Western Switzerland, 5210 Windisch, Switzerland\\
                   $^{5}$ Stuart D. Bale\\Physics Department, University of California, Berkeley, CA  94720, USA\\
                   $^{6}$ P. Higgins\\Astrophysics Research Group, School of Physics, Trinity College Dublin, Dublin 2, Ireland \\
                   email: \url{pohuigin@gmail.com}\\
                   $^{7}$ Y.~Liu\\State Key Laboratory of Space Weather, National Space Science Center, Chinese Academy of Sciences, Beijing 100190, China\\
                   email: \url{liuxying@spaceweather.ac.cn}\\
                    }

\begin{abstract}
On 13 March 2010 an unusually long duration event was observed by radio spectrographs onboard the STEREO-B and \textit{Wind} spacecraft. The event started at about 13:00 UT and ended at approximately 06:00 UT on 14 March. The event presents itself as slow drifting, quasi-continuous emission in a very narrow frequency interval, with an apparent frequency drift from about 625 kHz to approximately 425 kHz. Using the \inlinecite{springerlink:10.1023/A:1005049730506} interplanetary density model we determined that the drift velocities of the radio source are $\approx$33~km s$^{-1}$ and $\approx$52~km s$^{-1}$ for 0.2 and 0.5 times the densities of Leblanc model, respectively with a normalization density of 7.2~cm$^{-3}$ at 1AU and assuming harmonic emission. A joint analysis of the radio direction finding data, coronograph white-light observations and modeling revealed that the radio sources appear to be localized in regions of interaction with relatively high density and slow solar wind speed.
\end{abstract}

%
\keywords{Solar radio physics, Type II radio bursts}

\end{opening}

%
 \section{Introduction}

Type II radio bursts are common phenomena in coronal and interplanetary radio observations, identified as downward drifting features in frequency with source velocities much slower than those associated with type III radio bursts (\textit{e.g.} \opencite{1950AuSRA...3..387W}; \opencite{1985srph.book..333N}). It is generally accepted that type II bursts are caused by shock waves in the corona and CME-driven shocks in the solar wind. The energetic electron beams accelerated at the shock front excite Langmuir waves which through non-linear wave-wave interactions, couple into freely propagating radio emissions at the local electron plasma frequency ($f_{\mathrm{pe}}$) or its harmonic \cite{1982SoPh...78..187C,1989SoPh..120..393K,1999GeoRL..26.1573B}.

The radio emission depends on the local plasma density, $f_\mathrm{kHz} = a\sqrt{n\mathrm{(cm^{-3})}}$ where $a$ is 9 or 18 depending  on whether the emission is generated at the fundamental or harmonic frequencies, respectively. Therefore, the frequency of the radio emission indicates the heliospheric distance at which it was generated. In the interplanetary space, the plasma density, $n$, decreases with increasing heliospheric distance, $n(r) \approx 3.3\times10^5 r^{-2} + 4.1\times10^6 r^{-4} + 8.0\times10^7 r^{-6}$ cm$^{-3}$, for a  normalized electron density at 1 AU of $n_\mathrm{e}$(1 AU) = 7.2 cm$^{-3}$ \cite{springerlink:10.1023/A:1005049730506}. Thus, the type II radio emission is observed to drift down in frequency as the shock propagates outwards. Assuming a constant speed,  plotting the radio intensity as a function of inverse frequency ($1/f$) and time produces a straight line. This radio emission generated at coronal mass ejection (CME) shocks has been shown to be a useful tool to track CMEs through the interplanetary medium (\textit{e.g.} \opencite{1998JGR...10329651R}, \citeyear{1998GeoRL..25.2493R}; \opencite{2013ApJ...769...45L}).

The frequencies at which type IIs are generated are divided into at least three domains: metric, decameter-hectometric (DH) and kilometric. The WAVES experiment onboard the twin \textit{Solar TErrestrial RElations Observatory} spacecraft (STEREO A/B; \opencite{2008SSRv..136..487B}) covers emission in the DH and kilometric domains. STEREO/WAVES (SWAVES) includes two high frequency radio receivers (HFR1 and HFR2) with a combined observational frequency range between 0.125 and 16.075~MHz. SWAVES also has a low frequency receiver observing from 10~kHz (close to the electron local plasma frequency at 1 AU) to 160kHz. The SWAVES instrument therefore covers the DH domain, which corresponds approximately to heliospheric distances of 2 to 10 solar radii ($R_{\rm S}$, and the kilometric frequency domain, which corresponds to heliospheric distances beyond about 10$R_{\rm S}$. Similarly the Radio and Plasma Waves Experiment (WAVES); \cite{1995SSRv...71..231B} onboard the \textit{Wind} satellite covers the frequency range from 20~kHz to 14MHz.

It is generally accepted that kilometric type II radio emissions are generated by shock-accelerated electrons as the CME propagates in interplanetary space \cite{1999GeoRL..26.1573B,2005EOSTr..86..112V,Dauphin06}. However in the DH range the shock could be also produced by a flare (\textit{e.g.}  \opencite{Vrsnak06}; \opencite{2012ApJ...750...44B} and reference within).The relation between CMEs and type II emission has been extensively studied by many authors in the past (\textit{e.g.}  \opencite{2001JGR...10629219G}, \citeyear{2009SoPh..259..227G}; \opencite{2009ApJ...691L.151L}). Nevertheless, there are other possible scenarios for the generation of type-II like radio emission. One interesting scenario was proposed by  \inlinecite{2003ApJ...590..533R} to explain a long duration type II radio event that was observed for a period of about three hours. They suggested that the observed radio emission was generated in the regions of interaction between a coronal streamer and a CME-driven shock which propagated through the high-density streamer (see Figure~5 in \opencite{2003ApJ...590..533R} ). This results in an  enhancement of radio emission due to the increased density in the upstream medium which reduces the Alfv\'en speed, and increases the Mach number of the shock. \inlinecite{1998GeoRL..25.2493R} validated this scenario by  studying an unusual radio feature associated with an Earth-directed CME that occurred on 6 January 1996. They concluded that the location of the radio emission was well correlated with the position of a corotating interaction region (CIR). Although CIRs and streamers are different interplanetary structures, the physical processes involved are quite similar. 

In this paper we study the propagation and behavior of a slow drifting radio event detected by the WAVES instrument onboard the spacecraft STEREO-B and \textit{Wind}/WAVES on 13-14 March 2010. We track the radio emission in the interplanetary space using radio direction finding techniques and by fitting the observed radio spectrogram data.

\begin{figure} [htb]
\centerline{\includegraphics[trim=5mm 15mm 15mm 15mm,clip=true ,width=1\textwidth]{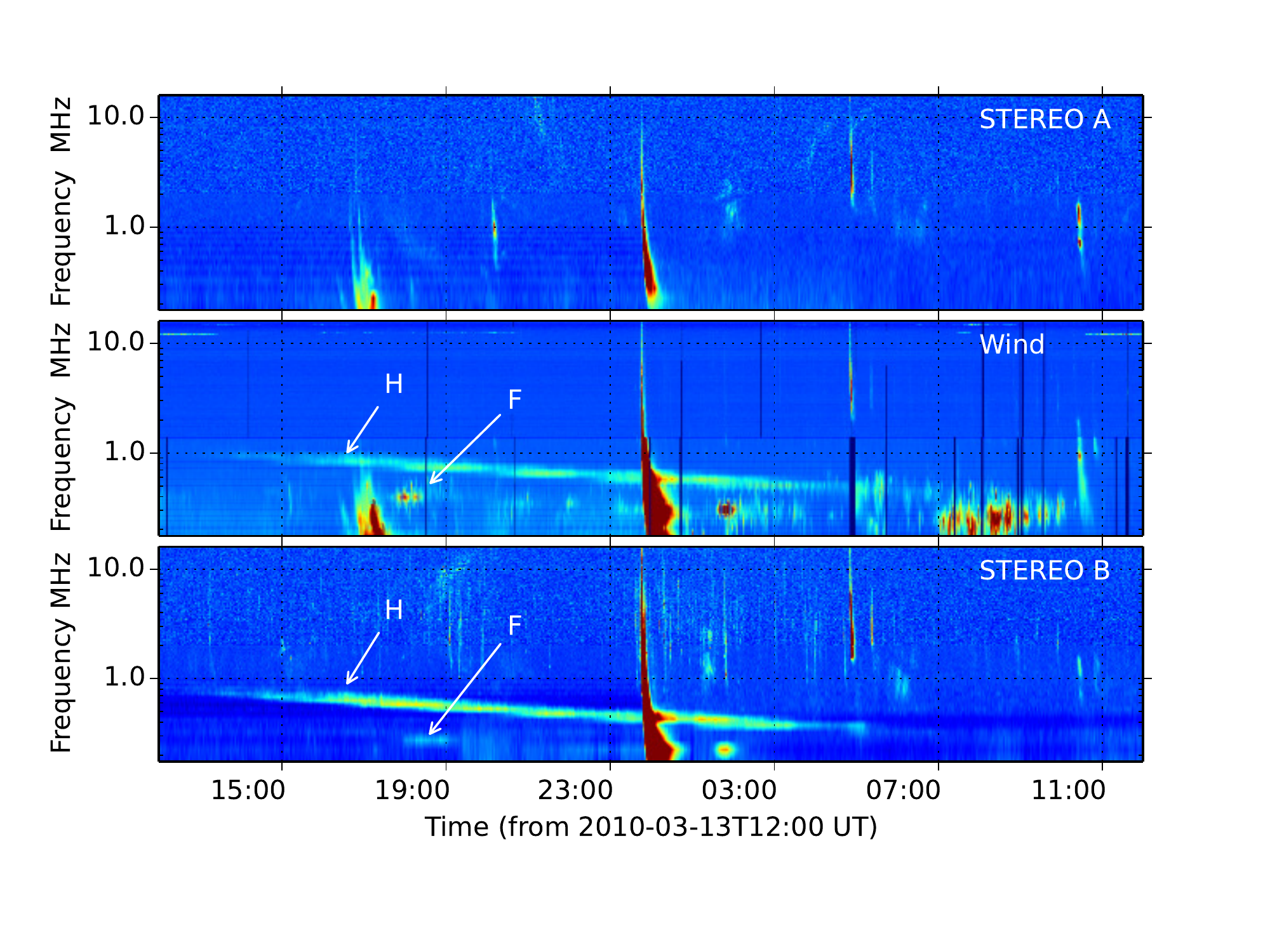}}
\caption{STEREO-A, STEREO-B, and \textit{Wind} dynamic spectra of the slow drifting event from the 13 March 2010 12:00 UT to 14 March 2010 12:00 UT. The plotted frequency range from 125 kHz to 16.025 MHz for STEREO and \textit{Wind}. The color shading represents the intensity of the radio emission measured in arbitrary units. The fundamental and harmonic emission are labeled F and H respectively.}
\label{fig1}
\end{figure}

 \section{Observations}
 \subsection{Radio Observations}
 
On 13 March 2010 an unusually long duration event was recorded by the radio instruments onboard STEREO-B and \textit{Wind}. The event started at about 13:00~UT on March 13 and ended at approximately 05:00~UT on 14 March. The radio source appears as a  slow drifting, quasi-continuous emission in a narrow interval of frequencies, ranging from about 625~kHz to approximately 425~kHz (Figure~\ref{fig1}), in what appears to be harmonic emission. 

The frequency drift rate of the burst, $df/dt$, is related to the radial velocity of the driving shock, $v_{R}$, via the relation: 

\begin{equation}
\frac{df}{dt} = v_R\frac{df}{dR}\\
\end{equation}

\noindent where $df/dR$ depends on the density model used, since the frequency of the emission is related to the density. To fit the burst we use the approach of \inlinecite{1997ESASP.415..183R}, where a linear fit is applied to the burst emission in plots of $1/f$ \cite{reiner07}. The frequency drift rate of the burst is related to the propagation speed of the shock, the gradient of the coronal electron density, and the angle of propagation with respect to the radial drop in $n_\mathrm{e}$. In order to determine the speed of the CME shock, a model for $n_\mathrm{e}$ is required. Using the \inlinecite{springerlink:10.1023/A:1005049730506} interplanetary density model we determined that the drift velocities of the radio source are $\approx$33~km s$^{-1}$ and $\approx$52~km s$^{-1}$ for 0.2 and 0.5 times the Leblanc model, respectively with a normalization density of 7.2~cm$^{-3}$ at 1AU and assuming harmonic emission. Table~\ref{table} shows the starting and ending height for the different folding factors. The folding factors were estimated from the proton density ($n_\mathrm{p} = n_\mathrm{e} + n_\alpha$) measured by the Plasma and Suprathermal Ion Composition (PLASTIC) on board the STEREO spacecraft. 

\begin{table}[htdp]
\begin{tabular}{c c c c}
Fold & $h_\mathrm{start}$  ($R_{\astrosun}$) & $h_\mathrm{end}$  ($R_{\astrosun}$)& Velocity (km s$^{-1}$)\\
\hline
0.2 &  7.21 & 12.75 & 33 \\
0.5 & 10.87 & 21.4  & 52 \\
\hline
\end{tabular}
\caption{Estimated radial distances to the Sun for the start and end of the radio emission and drift velocities derived from \protect\inlinecite{springerlink:10.1023/A:1005049730506} density model for 0.2 and 0.5 folding factors.}
\label{table}
\end{table}%

\subsection{Direction Finding}

Radio direction-finding (DF) is a technique that allows us to retrieve the direction of arrival and polarization state of incoming electromagnetic waves from radio measurements performed using a system of electric dipole antennas. There are different methods and techniques to retrieve the direction of arrival, depending on whether spacecraft is spinning or three-axis stabilized. \textit{Wind} is a spinning spacecraft and the direction of arrival of radio waves is determined by analysing the demodulated intensity of the electric field measurements and by simultaneously fitting these measurements to the equations for the antenna response during each spin period \cite{1972Sci...178..743F,1978JGR....83..616G,1985A&A...153..145F,1998JGR...103.1923R}. The result of this fitting process is the direction of the radio source, its polarization state and an estimate of the source angular size. STEREO is a three-axis stabilized spacecraft, for which several algorithms had been created to recover the directional information. We applied eigenvector \cite{Martinez11} and singular value decomposition \cite{santolik2003} algorithms to retrieve the directional information from the STEREO/WAVES data. STEREO/WAVES provide the auto- and cross-correlation voltages induced by the interacting with the antennas electromagnetic waves. The auto- and cross-correlations can be then organized in what is known as the \textit{spectral matrix}. From here the eigenvector is determined using for example a singular value decomposition, and the direction of arrival of the incoming electromagnetic wave can be calculated (for more details see \textit{e.g.} \opencite{2012JGRA..117.6101K}).

\begin{figure} [tb]
\centerline{\includegraphics[width=0.9\textwidth]{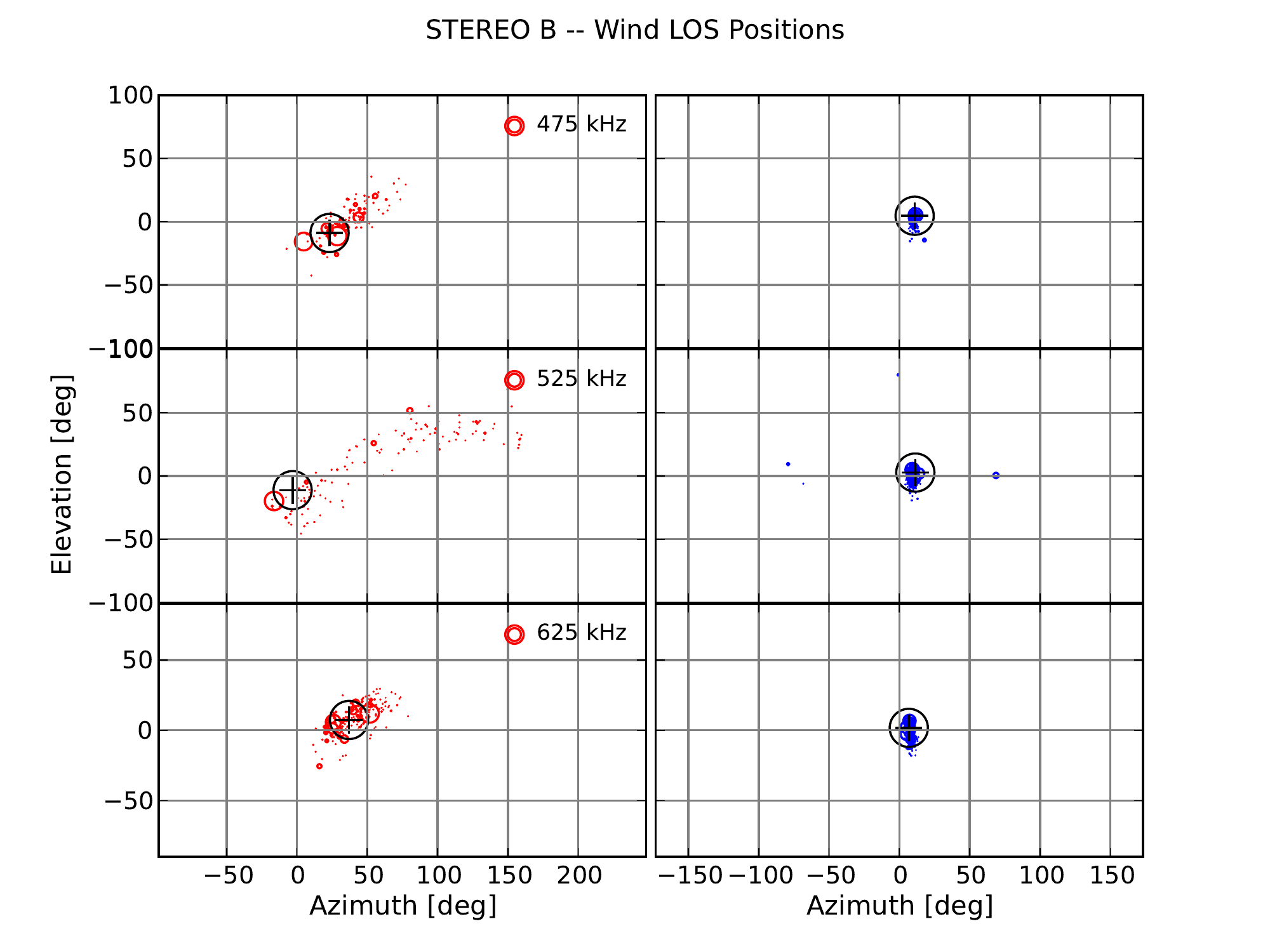}}
\caption{Derived line of sight positions of the radio sources for each of the two spacecraft, red for STEREO-B and blue for \textit{Wind}. The size of the circles corresponds to the intensity of the radio emission. The center of mass location is shown by crossed black circles whose radius does not have physical meaning.}
\label{fig2}
\end{figure}

The radio feature drifts between 625~kHz to 425~kHz and reaches a maximum intensity between 16:00~UT and 00:00~UT on 13 March 2010.  A type III burst was observed at about 23:50~UT on the same day. Two different techniques were use to compute the positions from STEREO-B, namely a singular value decomposition (SVD) \cite{santolik2003} and an eigenvalue decomposition \cite{Martinez11} method, obtaining similar results. Figure~\ref{fig2} shows the results obtained with the SVD decomposition algorithm (azimuth and elevation in the plane of the sky) for three frequencies: 475, 525, and 625~kHz; while for  \textit{Wind}/WAVES data the demodulation and fitting results were calculated for 484, 524, and 624~kHz. The apparent difference between the \textit{Wind} and STEREO-B derived positions is a consequence of the spacecraft different points of view in the heliosphere. During the period of observation STEREO-B had an angular separation with the Earth of about 71.5$^\circ$, and was located at a distance of 1.003AU from the Sun. \textit{Wind} was located at the $\mathrm{L_1}$ Lagrange point, meaning that its separation angle was close to zero and its distance to the Sun was about 0.985AU. 

\begin{figure} [!htbp]
\centering
\includegraphics[trim = 1mm 78mm 1mm 2mm,width=0.95\columnwidth]{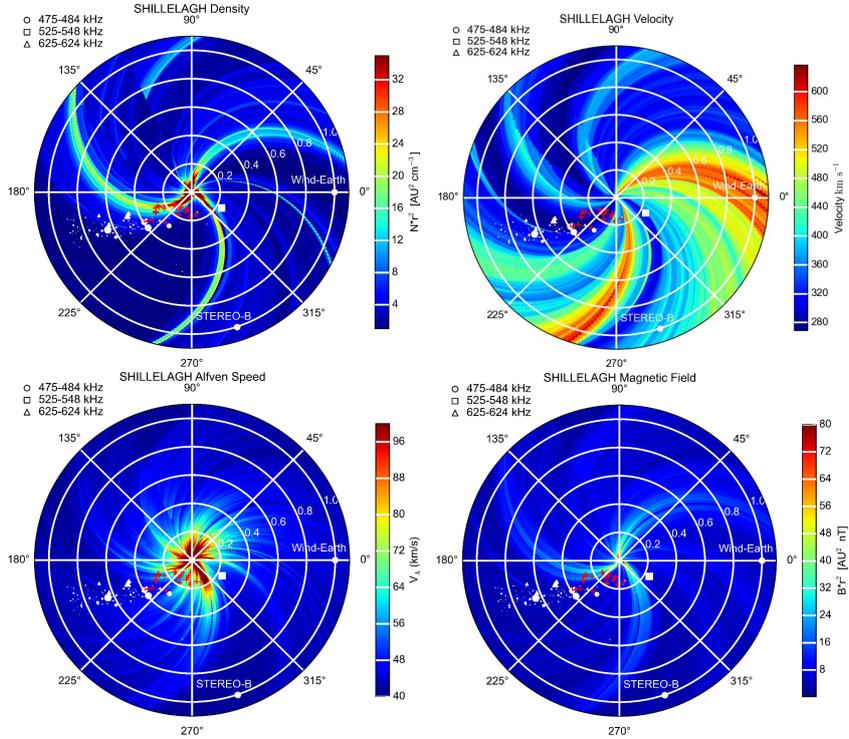}
\caption{Location of the geometrically triangulated positions of the radio sources in interplanetary space for three frequencies: 475--484 (circles), 525--524 (squares), and 625--624 (triangles)~kHz, as seen from the top. The size of the markers shows the normalized intensity of the radio emission. The triangulated positions of the CME launched at 14:54~UT is shown with red crosses. The background maps are four solar wind parameters (density, velocity, Alfv\'en speed and, magnetic field magnitude) calculated with the Solar wind-Heliosperic Imaging in Latitude and Longitude by Estimating Large-scale Attribute code (SHILLELAgh). These are results projected on the ecliptic plane and the size of the symbols represent the normalized intensity of the triangulated radio sources.}
\label{fig3}
\end{figure}

The triangulated position of the radio source, assuming that the source distribution is spherical, was computed using a two-dimensional (2-D) geometric triangulation algorithm based on that of  \inlinecite{2010ApJ...710L..82L}. This algorithm finds the location on the ecliptic plane where the two projected position vectors intersect, based on the line-of-sight azimuths from each spacecraft (See Figure~\ref{fig3}). These lines are the projections onto the ecliptic plane\footnote{The projections onto the ecliptic or equatorial plane are used in the triangulation process becuase in general the two vectors strictly do not intersect (\textit{e.g.} \opencite{Martinez11}, \opencite{2010ApJ...710L..82L}).} of the lines of sight between the individual spacecraft to the source as determined by the direction finding method. 

For the triangulation of radio sources the peak or average value of a number of azimuth and elevation measurements is typically used. Because the radio emission was produced over a long period of time, it is reasonable to assume that the radio source moved during this interval of time or that the radio emission was produced by multiple sources with timescales less than the  lifetime of the entire process. For this reason we did not use in our analysis a `unique' averaged position for each frequency, although we did calculate the center of mass as a sanity check (Figure~\ref{fig2}, black crossed circles). The position of the radio sources found with the two-dimensional triangulation technique throughout the duration of the burst with a cadence of $\approx$48 seconds are shown in Figure~\ref{fig3}, where the circles, squares and triangles represent three of the STEREO--\textit{Wind} frequencies used (475--484, 525--524, and 625--624~kHz respectively). The position of the radio sources are plotted on top of the density, solar wind speed, magnetic field, and Alfv\'en speed maps. The solar wind parameter maps were computed using the Solar wind-Heliosperic Imaging in Latitude and Longitude by Estimating Large-scale Attributes (SHILLELAgh\footnote{The SHILLELAgh software is available as a GIT repository: \url{https://github.com/pohuigin/SHILLELAGH}}) solar wind model (Suarez-Perez, D., Higgins, P., Gallagher, P.T., Krista, L., private communication).

\subsection{Ray Tracing Analysis}

As mentioned above, the positions obtained using direction-finding algorithms were used to recover the location of the bursts for time intervals throughout the burst via simple triangulation. It is well known that solar radio emission can be affected by propagation effects, in particular by refraction, having an impact on the measured directions and times of arrival. To understand how the local density will affect these positions we recomputed the position at the peak, taking into account refractive effects, via ray-tracing for two density profiles arbitrarily selected. We used the formalism presented in \inlinecite{2011ApJ...734...16T}: assuming that heliospheric densities are spherically symmetric, and using a density profile \cite{springerlink:10.1023/A:1005049730506}, ray-tracing by simple numerical integrations can be done (see \inlinecite{2011ApJ...734...16T} for details). The intersection of rays originating from each spacecraft is the location of the source. In case of the standard Leblanc atmosphere, differences in location with the non-refractive case were found to be about 0.05 AU. As a conceptual experiment, If 10 times the Leblanc densities are assumed, large differences of up to 0.5 AU are found.

\subsection{White-light Observations}
\label{cmewltr}

With the launch of STEREO it has been possible to observe and study the propagation of CMEs and other structures in the interplanetary space using white light observations ({\it \textit{e.g.}} \opencite{2008GeoRL..3510110R}; \opencite{2010GeoRL..3724103M}; \opencite{2010ApJ...710L..82L}, \citeyear{2013ApJ...769...45L}). We used this approach to determine if structures in the interplanetary space can be associated with the radio source positions obtained using direction finding. The different viewpoints of the STEREO spacecraft were exploited to identify the location of the radio emission, relative to the CMEs and other interplanetary structures (see Figure~\ref{fig3}). The heliographic coordinates of the CMEs front at different times were calculated by simultaneously selecting the same structures in both STEREO-A and -B images (tie-point method, \opencite{2011JASTP..73.1173L}) with specialized routines available in the SolarSoft package suite \cite{1998SoPh..182..497F}.

\begin{figure}[htbp]
\centering
\begin{minipage}{.49\columnwidth}
\includegraphics[trim = 20mm 0mm 20mm 0mm ,width=\columnwidth]{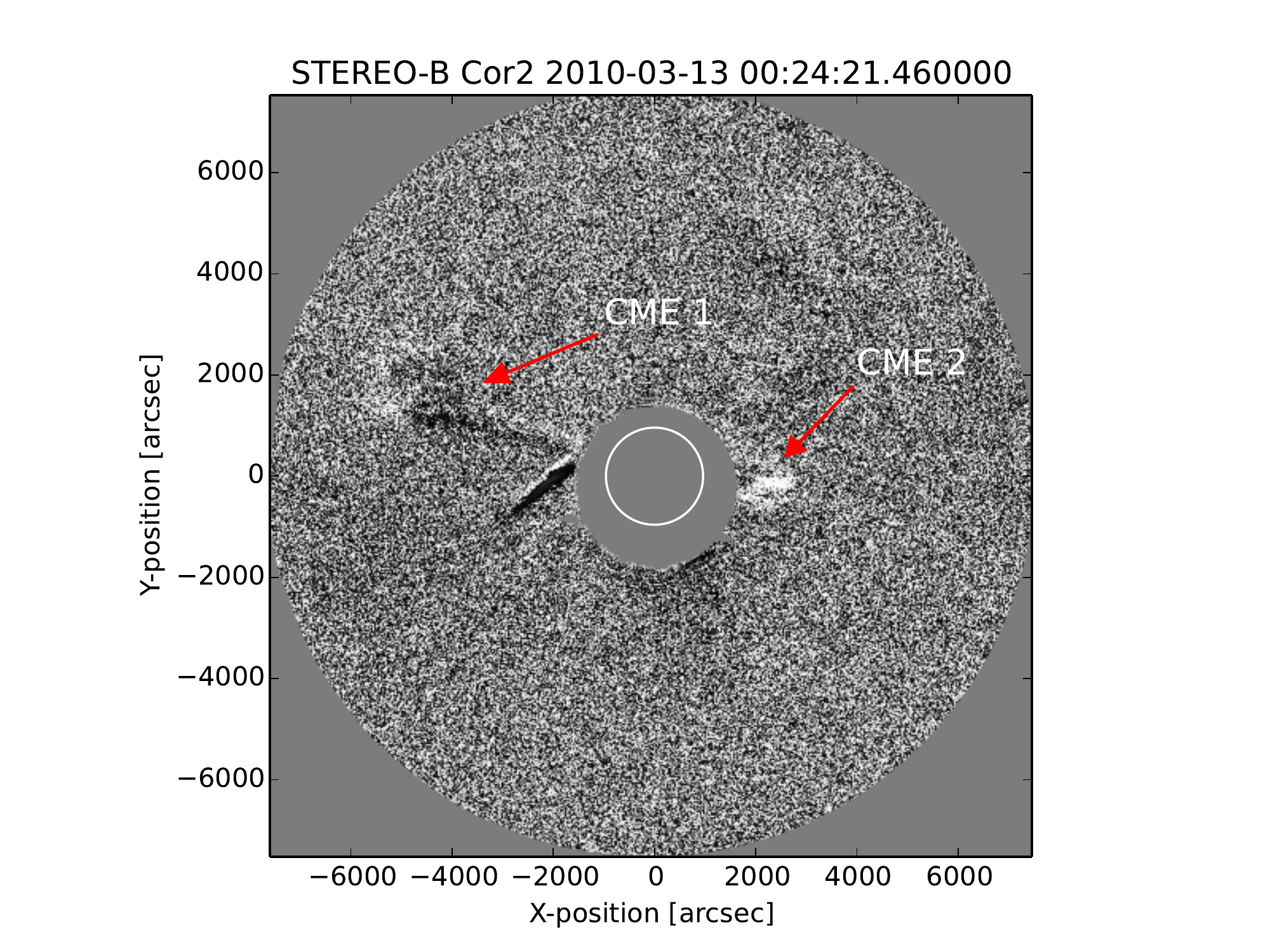}
\end{minipage}
\begin{minipage}{.49\columnwidth}
\includegraphics[trim = 20mm 0mm 20mm 0mm ,width=\columnwidth]{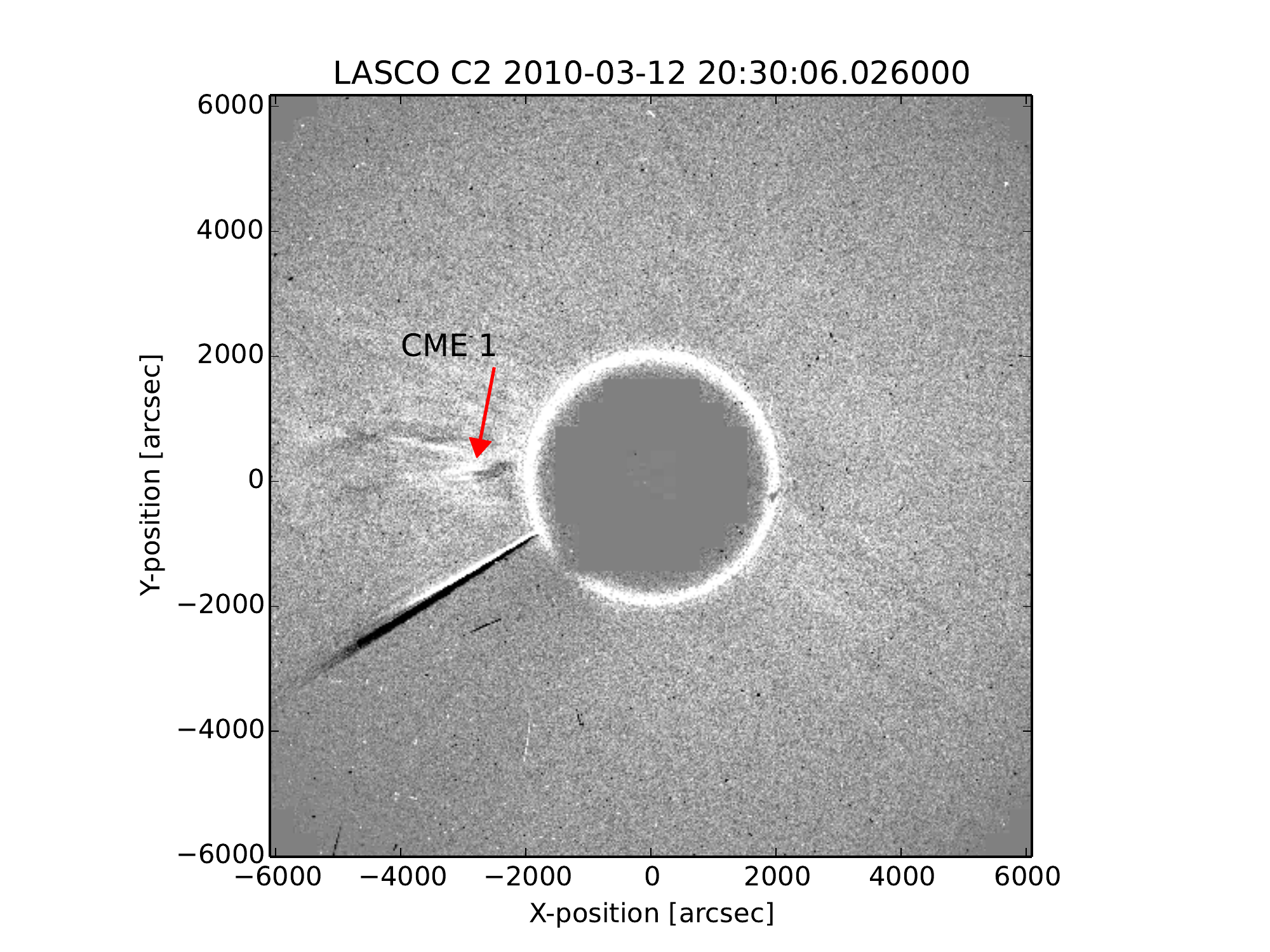}
\end{minipage}
\caption{White-light images of the two CMEs labeled as CME1 and CME2 as seen by the LASCO C2  (left) and STEREO-B Cor2 (right) instruments.}
\label{fig3.0}
\end{figure}

 \section{Discussion}
 
The 13 March 2010 radio event was observed by at least two spacecraft in the radio frequency spectrum. The event presented itself as a long duration slow drifting radio emission in a narrow interval of frequencies ($\approx$625\,--\,425 kHz). These kind of events have been observed before and the literature contains several suggestions to explain their nature. One of such hypothesis was proposed by \inlinecite{2003ApJ...590..533R}, who suggested that the observed long duration radio emission could be enhanced during the interaction between an expanding CME-driven shock and a streamer. 

The radio source positions were calculated using radio direction finding techniques which allow one to obtain the three dimensional locations of the radio source in the interplanetary space. By applying the method described by \inlinecite{2011ApJ...734...16T} we demonstrated that the triangulated positions of the radio sources are highly dependent on the local plasma density. Assuming that the local plasma density varies near the values predicted by the \inlinecite{springerlink:10.1023/A:1005049730506} model we conclude that the simple triangulation is a valid approximation, but in case that the density variation is higher this approximation is no longer valid and may lead to erroneous conclusions. Comparing the results of the radio direction finding with a model of the interplanetary space, by applying the SHILLELAgh solar wind model to in situ measurements, revealed that the radio sources appear to be localized in region of relative higher density and slow solar wind speed. The results of this comparison are shown in Figure~\ref{fig3}, where it is clear that the radio sources are located over an extended region of space and in close proximity to what appears to be a streamer. 

The idea that the radio emission is somehow facilitated by the interaction between a CME and a streamer certainly is appealing (see \opencite{1998GeoRL..25.2493R} and references therein). If one assumes the prior statement, as was proposed in \inlinecite{1998JGR...10329651R}, we estimated a CME lift-off time at about 22:00~UT on 12 March 2010. From the Cactus CME catalog \cite{2004A&A...425.1097R} we determined which CMEs were launched during the period of time that the slow drifting feature was observed. Cactus reported two possible CMEs, one launched on 12 March 2010 at about 14:54~UT (CME1 in Figure~\ref{fig3.0}), and a second at approximately 21:54~UT (CME2). The second coronal mass ejection seems to fit the prediction made by our initial analysis of the radio spectra, however we were not able to determine the three dimensional position of the CME, as this event was very faint and only observed by cameras on STEREO-B and the propagation direction does not correspond with the derived radio positions, except for one period of time. The first CME seems to match the direction of the radio source, although the predicted liftoff time seems early. The velocity of the CME computed by CACTus was $\mathrm{301 \pm 56}$~km s$^{-1}$, which is about six times the value obtained by our fitting of the radio spectra for a folding factor of 0.5. So, it seems plausible the existence of certain relation between the radio emission and a possible interaction between this faint CME and a streamer. The fact that we could not determine the position of the CME prohibits us from making a conclusive statement. The triangulated position for the second CME was possible to determine using the technique presented in Section~\ref{cmewltr}, showing that the general direction of propagation of this CME is align with the calculated radio source positions. The interaction between the CME-driven shock and the steamer seems like a good candidate to explain the observed radio emission, in which the CME-driven shock gets ``reenergized"  as suggested by Reiner (1998), and becomes visible in the radio spectra at approximately 13:00~UT on 13 March 2010.  

This is a very challenging observational event and it is clear that a more detailed statistical study must be done, by compiling a larger set of this type of event during the operational time of the STEREO spacecraft. This future study should be accompanied by a theoretical analysis that may shed light on the physical processes responsible for this radio emission. Future missions out of the ecliptic may contribute to a more accurate determination of the CME positions, and therefore a better understanding of the physical conditions that lead to the radio emission. Missions like \textit{Solar Probe Plus} and \textit{Solar Orbiter} will prove invaluable for the measurement and generation of more accurate solar wind and density models that in turn will corroborate where and how the radio observed slow drifting radio emissions (like the one presented in this article) were generated.
 
\acknowledgements
This work was supported by NASA under Contract NNX12AG98G for authors Mart\'inez Oliveros, Bain, Raftery and Krucker. V. Krupar acknowledges the support of the Czech Grant Agency grant 205-10/2279. This research has made use of SunPy, an open-source and free community-developed solar data analysis package written in Python \cite{mumford-proc-scipy-2013}. 

 \bibliographystyle{spr-mp-sola}
 \bibliography{MyLibrary}

\end{article} 

\end{document}